# High-contrast imaging of tight resolved binaries with two vector vortex coronagraphs in cascade with the Palomar SDC instrument

Jonas Kühn,*[a] Sebastian Daemgen,[a] Ji Wang,[b] Farisa Morales,[c] Michael Bottom,[c] Eugene Serabyn,[c] Jean C. Shelton,[c] Jacques-Robert Delorme,[b] and Samaporn Tinyanont[b]

[a]Institute of Particle Physics and Astrophysics, ETH Zurich, Wolfgang-Pauli-Strasse 27, 8093 Zurich, Switzerland
[b]Department of Astronomy, California Institute of Technology, 1200 E. California Blvd., Pasadena, CA 91125, USA
[c]Jet Propulsion Laboratory, California Institute of Technology, 4800 Oak Grove Dr., Pasadena, CA 91109, USA

**ABSTRACT**

More than half of the stars in the solar neighborhood reside in binary/multiple stellar systems, and recent studies suggest that gas giant planets may be more abundant around binaries than single stars. Yet, these multiple systems are usually overlooked or discarded in most direct imaging surveys, as they prove difficult to image at high-contrast using coronagraphs. This is particularly the case for compact binaries (less than 1'' angular separation) with similar stellar magnitudes, where no existing coronagraph can provide high-contrast regime. Here we present preliminary results of an on-going Palomar pilot survey searching for low-mass companions around ~15 young "challenging" binary systems, with angular separation as close as 0''3 and near-equal K-band magnitudes. We use the Stellar Double Coronagraph (SDC) instrument on the 200-inch Telescope in a modified optical configuration, making it possible to align any targeted binary system behind two vector vortex coronagraphs in cascade. This approach is uniquely possible at Palomar, thanks to the absence of sky rotation combined with the availability of an extreme AO system, and the number of intermediate focal-planes provided by the SDC instrument. Finally, we expose our current data reduction strategy, and we attempt to quantify the exact contrast gain parameter space of our approach, based on our latest observing runs.

## 1. INTRODUCTION

Almost half of all solar-type stars in our Milky Way are bound in multiple systems.[1] Yet, sub-stellar and planetary companions have mostly been probed around single stars, as they are easier to observe and model. Clearly, to obtain a complete understanding of planetary origins and diversity, we must consider binary stars because they are a common outcome of star formation. Currently there are >100 planets known to reside in binary stars[2] despite extreme dynamical conditions and a bias against their inclusion in previous surveys. This demonstrates that planet formation is a universal and robust process. As with single stars, most binary star planets have been discovered with the radial velocity technique or transit photometry. For these systems, no detailed characterization is possible because the planetary light cannot be separated from the host star light with high fidelity. On the other hand, direct imaging enables to obtain information about the luminosity of the planet, which may provide clues about its formation path when compared to evolution models. Furthermore, atmospheric composition may then be assessed with high signal-to-noise spectroscopy follow-up, making these observed planets ideal templates for comparison with other low-mass young and field objects.

*jonas.kuehn@phys.ethz.ch

Despite their exclusion from surveys, several binaries have been reported to host directly-imaged companions (ROXS42B, HD131399, 51 Eri).[3-5] This suggests that planetary companions may be frequent, possibly even more frequent than planets in single systems (<4.1% for FGK stars and orbital radii of 30–300AU;[6] cf. also Ngo et al. 2016 who report a 2.9-fold increase in multiplicity fraction for hosts of close-in hot Jupiter planets).[7] As these discoveries mostly stem from surveys that did not systematically explore binaries, there is currently no precise estimate for planet occurrence in binary star systems. Accordingly, new dedicated imaging observations would be needed to constrain planet occurrence in binaries. The expected return would be a unique chance of low-mass companion discovery, within a yet uncharted observational parameter space, as well as better statistics on companions in stellar multiples, even in the case of a null result. We would also expect to learn more about the planet formation process in general, by probing whether planets form and survive in dynamically active systems with a turbulent formation history.

To maximize the planet discovery potential, imaging surveys of single stars use adaptive optics (AO), coronagraphs, and high-level post-processing algorithms (e.g. ADI, LOCI),[8,9] to improve contrast close to the star by a few orders of magnitude. Dedicated direct imaging survey of binary stars have been recently undertaken (e.g., the VIBES and SPOTS surveys),[10,11] however there only one of the two stars – the brightest primary component - is nulled by a focal-plane coronagraph. To avoid saturation of the secondary star, short exposure times are required, which increase the noise level and limit achievable contrast. Furthermore, smearing and uncontrolled diffraction of the binary companion make the application of ADI methods less efficient, especially for angular separation of less 0''.5 from the primary. Even alternative dedicated strategies with no coronagraph, such as Binary Differential Imaging (BDI),[12] quickly degrades for close binaries and for systems with magnitude ratios <2 mag.

## 2. METHODOLOGY

### 2.1 The Palomar SDC instrument: A unique platform for high-contrast imaging of binary stars

As previously mentioned, multiple-hosting star systems are often excluded from large high-contrast imaging (HCI) surveys,[13-15] because they prove difficult to observe with focal-plane coronagraphs as the off-axis companion star(s) is leaking through, being unaffected by the coronagraph suppressing the primary. This is especially true for small brightness ratio systems at close separation (less than 1''). In principle, pupil-plane coronagraphs (e.g. the Apodizer Phase Plate[16]) may be used to observe these systems, but are limited in terms of accessible field-of-view (FOV) and inner-working angle (IWA). Furthermore, pupil-plane coronagraphs can lead to potentially damaging unwanted saturation of the scientific detector outside the high-contrast FOV. We would therefore be looking for an ideal HCI instrument to observe binaries on a large telescope, consisting in an adaptive pair of small-IWA focal-plane coronagraphs that can be moved around to match a particular astrometric binary configuration, in combination with a state-of-the-art 2$^{nd}$-generation extreme AO system. Quite remarkably in the current HCI landscape, the Palomar Stellar Double Coronagraph (SDC) instrument[17,18] equipping the 200-inch Hale Telescope stands out by being able to combine several of such features required for coronagraphic imaging of binary stars.

SDC was originally conceived to improve the IWA and contrast of vector vortex coronagraphs (VVC)-assisted direct imaging, in presence of the non-ideal pupil that represents the Hale Telescope secondary mirror central obscuration.[19] Conceptually, the idea was to take advantage of the light-folding properties of vortex coronagraphs, using two of them on-axis in series to fold the diffracted light from the secondary back inside the central obscuration region of the pupil, where it can be masked most efficiently with no penalty on IWA and throughput, as successfully demonstrated on-sky.[17,20] To this purpose, the SDC instrument integrates two VVCs in cascade, based on liquid crystal Ks-band waveplates,[21] as shown of the optical layout cartoon of Figure 1.[18] SDC also provides two intermediate pupil planes (one transmissive and one reflective, see Fig.1) to insert Lyot stops or apodizing optics. The instrument is mounted downstream of the PALM-3000 (P3K) extreme AO system,[22] which integrates a 64x64 high-order deformable mirror from Xinetics (Fig.2a), running at up to 2 kHz. In practice, the P3K AO system can deliver best-in-class atmospheric correction (Strehl ratio above 0.8 for K-magnitudes of less than ~8), despite the relatively modest seeing conditions on Mt. Palomar (median seeing between 1'' and 1''.5). As shown on Figure 1, the SDC module is designed as an intermediate coronagraphic stage, and the imaging is done downstream of its output beam path, with the Palomar High Angular Resolution Observer (PHARO) NIR imager (Fig.2b).[23]

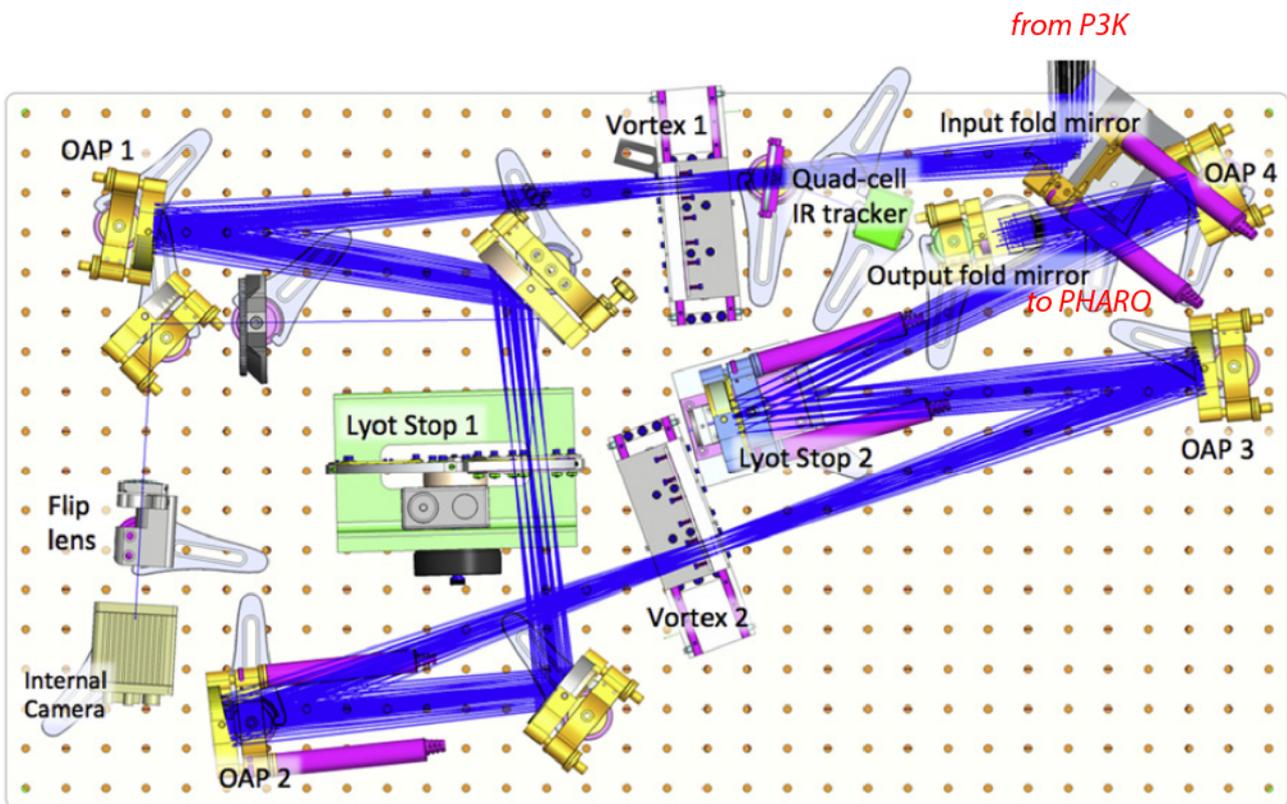

Figure 1. Optical layout of the Palomar Stellar Double Coronagraph (SDC) visitor instrument.[18] The near-infrared (NIR) portion of the spectrum is transmitted to the instrument from the P3K AO system as a converging beam (see top of Fig.). Following a first actuated fold mirror, the J-band spectra component is reflected towards a tip/tilt tracker (quad cell), whereas the H-K-band beam propagates through a succession of two $K_s$-band vortex coronagraphic stages (H-band vortices also available, but not used) and Lyot pupil planes. It is then folded "behind" the instrument, towards the PHARO NIR imager (not shown). Of particular interest for the observation of binary stars is the 2nd off-axis parabolic mirror (OAP 2), which can be remotely actuate to move the science-plane image laterally, in-between the two coronagraphic stages, hence making it possible to null two stars at the same time.

In the context of imaging binary stars, we first note that no opto-mechanical constrains force us to use both SDC VVCs on-axis, as the off-axis parabola (OAP 2, see Fig.1) focusing the light on the 2nd VVC (vortex #2) can be remotely mechanically actuated in all three degrees of freedom, and can therefore be used to align the second star to the center of vortex #2. In practice we will use SDC in an alternative configuration, by relying on the first stage of the instrument (vortex #1, its dedicated tip/tilt tracker and intermediate Lyot stop 1, see Fig.1) in the standard way, pointing the primary star behind vortex #1. In a second step, we can remotely tilt OAP 2 upstream of vortex #2 to move the secondary star such that it is suppressed by the vortex #2. Finally, we also note that the Hale Telescope is the world-largest telescope equipping an equatorial mount (see Fig.2c), resulting in the absence of sky rotation in the image-plane while observing. This important – and quite unique – property makes it possible to maintain any observed binary targets mechanically aligned between the pair of VVCs, once the alignment is completed. Indeed, a more classical VLT-class Alt-Az telescope would require operating in field-tracking mode, which would in turn make the pupil rotate and compromise the masking of the secondary support structure (spiders) in the Lyot pupil planes downstream of the coronagraphs. Overall, this unique combination of features likely makes the Palomar SDC instrument an ideal platform to observe binary systems, including challenging targets with close separations (less than 1'') and small brightness ratios (less than 3 mag difference at Ks-band).

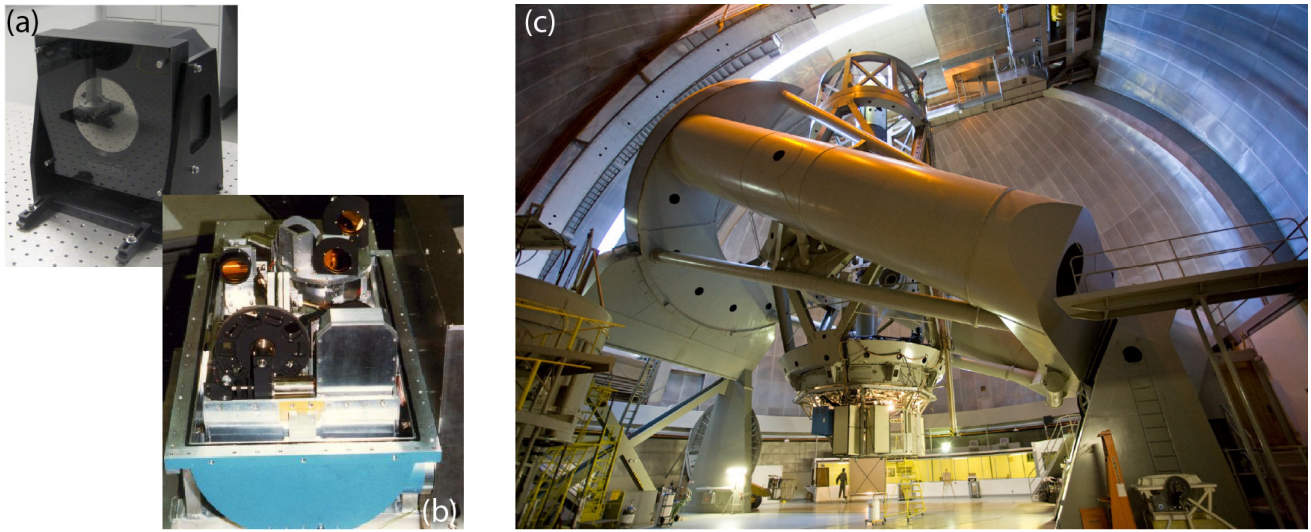

Figure 2. The unique capabilities of the Palomar 200-inch telescope, which – when combined with the SDC instrument – makes it strikingly suitable to observe binary stars. (a) The P3K extreme AO system[22] integrates a 64x64 DM from Xinetics, running at up to 2 kHz. (b) The PHARO instrument[23] provides y-K band imaging and slit spectroscopy capabilities, and can be mounted downstream of SDC. (c) The 200-inch telescope operates the world largest equatorial mount, hence there is no sky rotation when observing.

## 2.2 Coronagraphic imaging of binaries with SDC: Observational strategy

The absence of sky rotation on the 200-inch (Fig.2c) prevents the use of Angular Differential Imaging (ADI)[8,9] to improve PSF subtraction when reducing the coronagraphic data, hence the usual observational for high-contrast imaging at Palomar relies on so-called Reference star Differential Imaging (RDI). With RDI, one needs to interleave calibrator star acquisitions in-between science exposures, every 300 s or so, by quickly slewing the telescope to a reference star with similar sky location, spectral type and brightness as compared to the science target.[20] When observing binary stars, the situation obviously gets more complicated, and one needs to observe the reference star (same spectral type as the science primary) at both VVC locations, in order to get relevant speckle field data for both focal-plane positions (see §2.3). In terms of observational efficiency, this state of affair has a direct impact on the time-on-target and duty-cycle when observing a scientific target of interest, even more so when taking into account the initial delay to setup the pair of VVCs for a particular binary system.

As illustrated in Figure 3, when using SDC to image binary stars, the observing sequence for a given science target can be essentially divided into two phases: initial coronagraphic setup, and integration. During the initial setup, once the AO loop is closed on the primary (for equal-brightness binaries, the P3K AO remains stable for angular separation not less than 0''.2), one can remotely send AO "science" tip/tilt commands (centroid offsets) to point the primary star behind vortex #1. This is mostly done manually but, once this alignment is completed, one can use SDC first-stage J-band tip-tilt sensor (see Fig.1) to lock the entrance beam pointing. The coronagraphic alignment of the secondary star is then done is a semi-automatic way (Fig.3): the operator measures the RA-DEC image-plane coordinates (in pixels) of both stellar components on the PHARO image, and this information is then fed to a coordinate-transformation routine to generate approximate (typically accurate to within 1 $\lambda/D$) tip/tilt offset values, corresponding to the differential pointing required to move the image so that the secondary star is masked by vortex #2. To mitigate losses in wavefront quality, the X-Y mechanical move is achieved by a combination of Y-tilt of OAP 2 (Fig.1) and X-translation of the motorized optical stage holding vortex #2. Once this process is completed (it takes about 15 mn for a given binary target), one can start the integration phase. As shown in Figure 3, PHARO duty cycle is typically 50%, and one also needs to slew to a calibrator star at least every 300 s to get a pair of RDI integration sequences behind each vortex locations (typically half the length of a science sequence for each, hence totaling about the same time as a scientific exposure overall). In practice, the RDI sequences need to be undertaken with both VVCs still inserted in the beam train, and at the exact same position as for the scientific exposures. Overall, also adding non-coronagraphic data acquisitions (for photometry and astrometry) and sky background calibrations, plus initial AO setup time and contingencies, a typical 900-s integration-on-target data stack in the context of a survey ends up representing up to 1.5 of time-on-target. Hence the median yield per night is about 4-5 targets per night, depending on weather conditions and AO corrections stability.

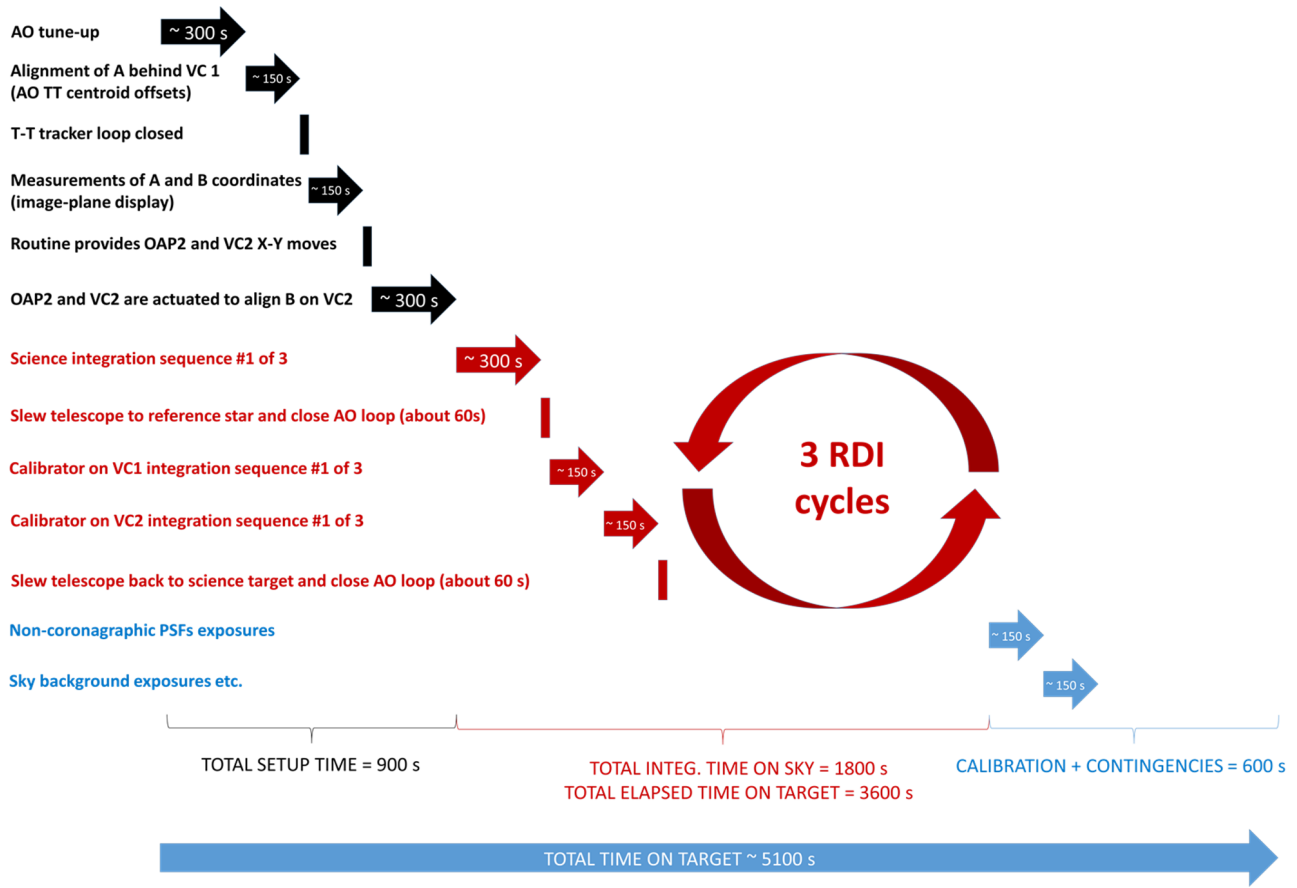

Figure 3. Flowchart of a typical observing sequence for a binary target with SDC on the 200-inch. Setup time for AO and SDC is in the order of 900 s, and 3,600 s are dedicated to science and reference star exposure (hence 1,800s of integration time given PHARO 50% overhead, about equally split between science and RDI PSF sequences). Taking a 600-s margin for calibration data (photometry, sky background) and contingencies, this ends up to about 5,100 s (1.4 hour) of time on target.

## 2.3 Data reduction approach

The retained preliminary data reduction strategy here is a naïve extension of the classical high-contrast RDI pipeline used to process PHARO or regular (non-binary) SDC vortex coronagraphy data.[20] As schematically illustrated in Figure 4, in a first phase the three data stacks – the science target sequence and the two calibrator star data stacks behind vortex #1, respectively vortex #2 – are treated separately for pre-processing. The latter is done on a frame by frame basis, and includes flat-fielding, sky background subtraction, bad pixel and cosmic ray event corrections, and frames registration (alignment), using the non-coronagraphic PSF exposures to determine centroids by aperture photometry. Both calibrator sequences are then concatenated in a "super cube", and a LOCI/PCA algorithm[24,25] is applied to generate principal components (PCs) pseudo-images. Given that these PCs then describe both calibrator star coronagraphic PSFs for each VVC, they encompass an accurate representation of both speckle fields that are incoherently superimposed on the science binary coronagraphic data stack. In this regard, it is important to acquire calibration RDI data with at least the same, and ideally slightly superior, photo-electron counts as compared to the science sequence. As a last step (see Fig.4), each individual science frame is projected onto the obtained PCs, and this PCA model of the instantaneous PSF is subtracted for the said frame. Finally, each reduced frame is adjusted for photometry using the non-coronagraphic PSF data (see Fig.4), and all of them are ultimately median-combined to produce the final reduced image

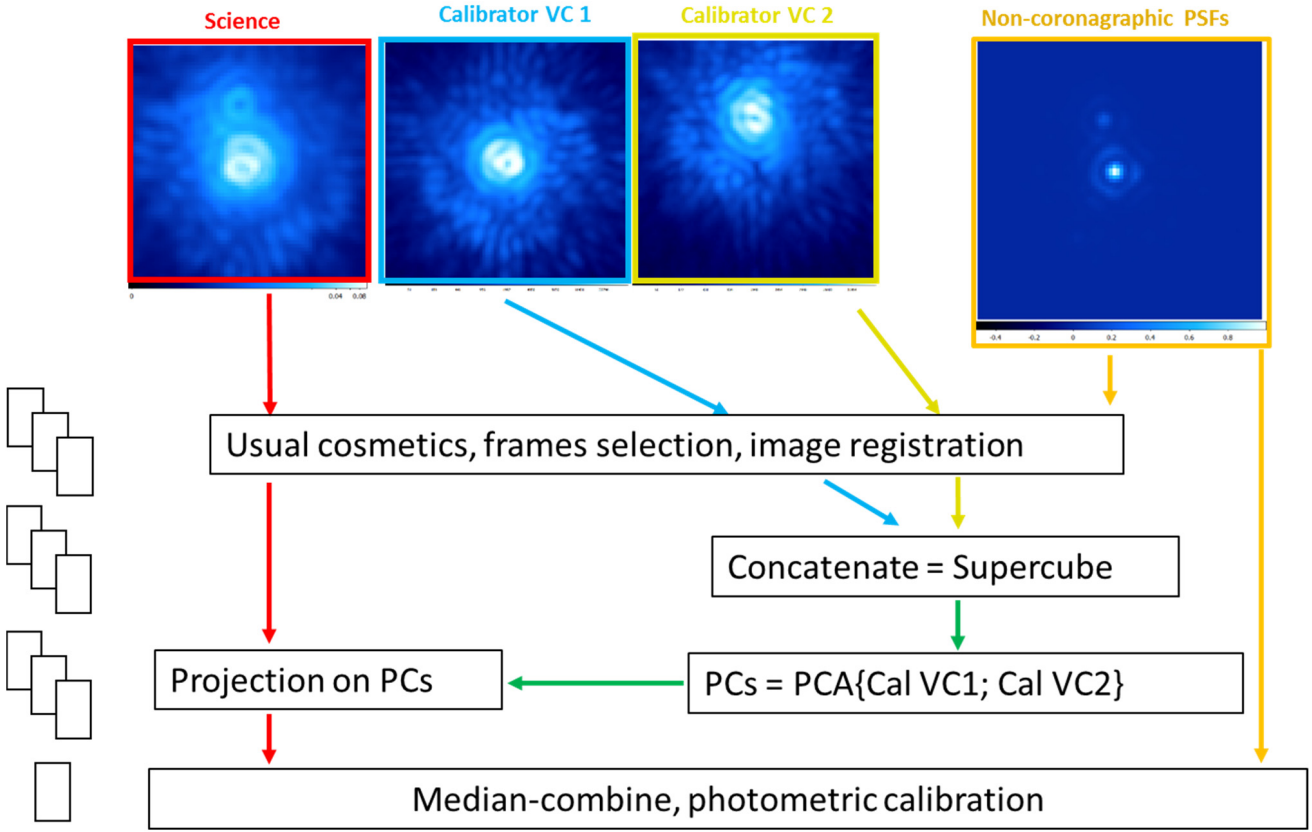

Figure 4. Cartoon of the retained approach for reducing SDC binary star data with RDI PCA. The calibrator star data cubes are concatenated together before computing the PCs, on which the science data will then be projected. The non-coronagraphic PSF data serves two purposes: astrometric, to align the reference star frames to the correct RA-DEC centroids, and photometric, for the final reduced data product.

## 3. PRELIMINARY RESULTS

### 3.1 Typical on-sky instrumental performance

We initiated a test campaign of 3 nights in February 2017, to verify the feasibility of instrumental approach described in sections §2.1 and §2.2. Based on the positive results, which are briefly exposed hereafter, we then applied for more telescope time on 200-inch each semester to conduct an exploratory survey of young binary stars (see §3.2). Here we give a brief overview of "typical" performance when observing targets in the context of such a small survey, with approximate total integration time on target of $\sim 900$ s for a total time on target of $\sim 7,200$ s, and this under an average seeing of $\sim 1''.3$. Figure 5 presents a typical on-sky example of the data reduction products described in section §2.3, in this case for the young triple system BD+005017 ($sep_{AB} \sim 1''.2$, $\Delta K_{AB} \sim 1.8$, $sep_{BC} \sim 0''.4$, $\Delta K_{BC} \sim 0.8$, Carina moving group). As can be seen on Fig.5c, the RDI PCA implementation of §2.3 enables to gain about a factor two in signal-to-noise (S/N) on the tertiary component, as compared to the median combined coronagraphic PSF (Fig.5b). Another important practical point is that – at least for sequences of less than 2-hours on target – we did not notice any significant differential tip/tilt inside

SDC, between the entrance J-band tip/tilt tracker (Fig.1) and the location of vortex #2, thus the initial SDC alignment setup (Fig.4) remains in place throughout the scientific exposures.

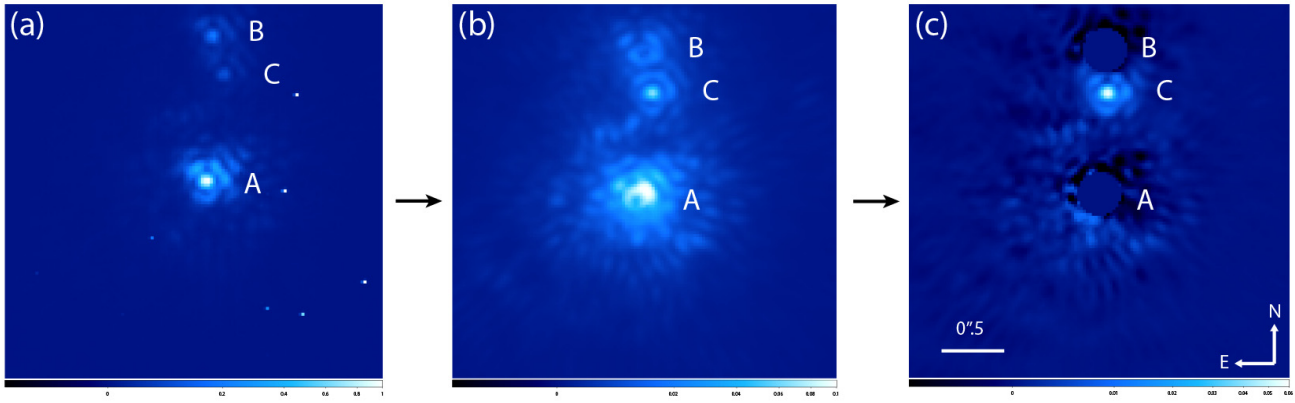

Figure 5. On-sky example of $K_s$-band vortex coronagraphy of binary stars using SDC, here on the triple system BD+005017 ABC. (a) Non-coronagraphic median PSF, with the 3 stellar components clearly visible. (b) Median-combined raw coronagraphic image, with the C component detected with a S/N of ~20. (c) RDI-PCA final reduced imaged, with the C component detected at a S/N of ~41. All images are stretched in asinh scale for improved visibility.

In terms of achievable contrast and detection limits of low-mass companions, Figure 6a shows a typical 5-sigma contrast curve – but under sub-optimal seeing conditions of ~1''.5 - after RDI PCA reduction as described in section §2.3, for a bright ($K_A$ ~ 4.5) binary, with a $\Delta K$ ~ 2 secondary component at ~ 0''.3 from the primary. No speckle nulling techniques[26] were used here, as it would further impact the observing efficiency (Fig.3), but those might be valuable to investigate for follow-up observations of candidates in the near future. Also shown on Figure 6a with the second vertical axis are the corresponding mass sensitivity limits in $M_{Jup}$, based on evolutionary models.[27] Figure 6b presents the derived theoretical detection limits in function of total usable integration time on target, assuming an optimistic scenario of photon-noise limited contrast (which is mostly only true beyond ~1''.5).

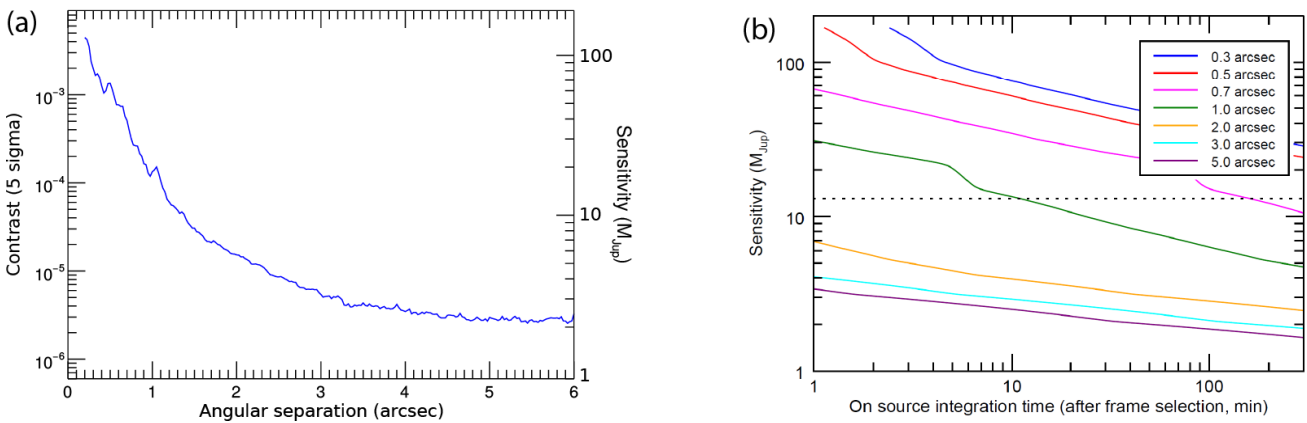

Figure 6. (a) Typical 5-sigma contrast curve and companions mass sensitivity upper limits for a bright (Kmag ~4.5) primary, with $\Delta K$ ~ 2 secondary component at ~ 0''.3, under 1''.5 seeing conditions and with an above-average usable integration time (after frame selection) on the science target of ~900 s. (b) Extrapolated detectable mass sensitivities curves in function of on-source integration time, based on (a) and assuming an optimistic scenario of contrast depending on photon statistics.

## 3.2 An attempt at estimating the contrast gain when observing binaries

An obvious question is when does it become valuable to use our proposed dual-coronagraph scheme to observe a particular binary star, as compared to a relying on the classical singe-coronagraph configuration? This is a complex question, as one can expect the contrast gain – or absence of gain - to depends on several target-related parameters, among which brightness ratio, projected angular separation, and absolute brightness of the primary (AO natural guide star), as well as more random environmental parameters like seeing, cloud transparency and wind conditions. We therefore here only attempt to provide one example of comparison between both coronagraphic setup, based on engineering data from December 2017, where we observed one particularly challenging binary (sep ~ 0''.3, DK ~ 1) with both configurations (one VVC or two VVCs, in an interleaved fashion to mitigate influence of environmental changes), and with similarly deep (~ 900 s) integration times on target. The results are presented in Figure 7, where the advantage of using two VVCs in cascade to observe this target is clear, with about one stellar magnitude contrast gain in the simple median-combined case, whereas the gain reaches about two stellar magnitudes using PCA. In this regard, it is striking to note that PCA struggles to improve contrast when the secondary star is not nulled.

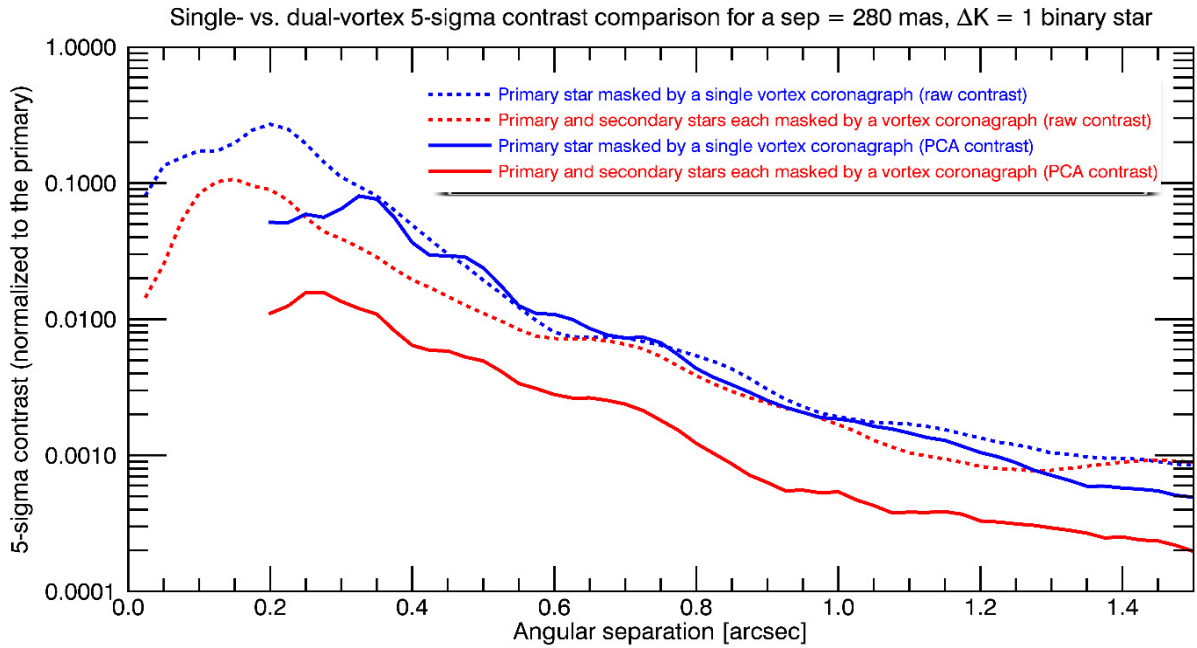

Figure 7. Contrast curves (5-sigma) for the same binary target, observed with two or only one vortex coronagraph (in the latter case only the primary is nulled), during the same night and in an interleaved fashion (to average out changes in observing conditions). Both raw (simple median-combine) and PCA final contrast curves are shown.

## 3.3 A pilot-survey of nearby young binary stars: Overview and current completion status

Since February 2017, we initiated a pilot survey of ~15 binary stars with separations ranging mostly between 0.3'' and 2''.[28,29] The objects are young (~1–500Myr), which serves two purposes: firstly, low-mass objects are brighter when young – and thus easier to detect – regardless of the formation process. Predictions of the minimum detectable mass in the proposed survey range from ~10MJup to ~2MJup (example shown in Fig.6a), depending on the host-star brightness, distance, and age (based on evolutionary models).[27] Secondly, the difference between two popular formation scenarios - "hot-start" and "cold-start"[30,31] - is largest at young ages. Here we expect to be sensitive to both planets on "stellar type" (s-type) orbits, orbiting one individual star of a binary, or wide "planetary type" (p-type) orbits around the center of mass of a close binary. The compelling possibility of observing faint circumbinary disk material is also real with the proposed scheme. The proposed observations would thus contribute to identify the most likely formation scenario for massive planets in wide orbits.

On the course of the period ranging from February 2017 to April 2018 – with 6 "usable nights" for 11 allocated nights, due to weather - we already completed 82% of our pilot survey (75% in terms of reduced data). A mosaic exposing a subset of our currently obtained results is exposed in Figure 8. We have two upcoming SDC observing nights on September 27 and 28, 2018, which should cover bring completion to 100%. At the time of writing, this pilot survey did not yield any planetary-mass companion or proto-planetary disk candidates, but two stellar-mass candidates remain to be followed-up for confirmation.

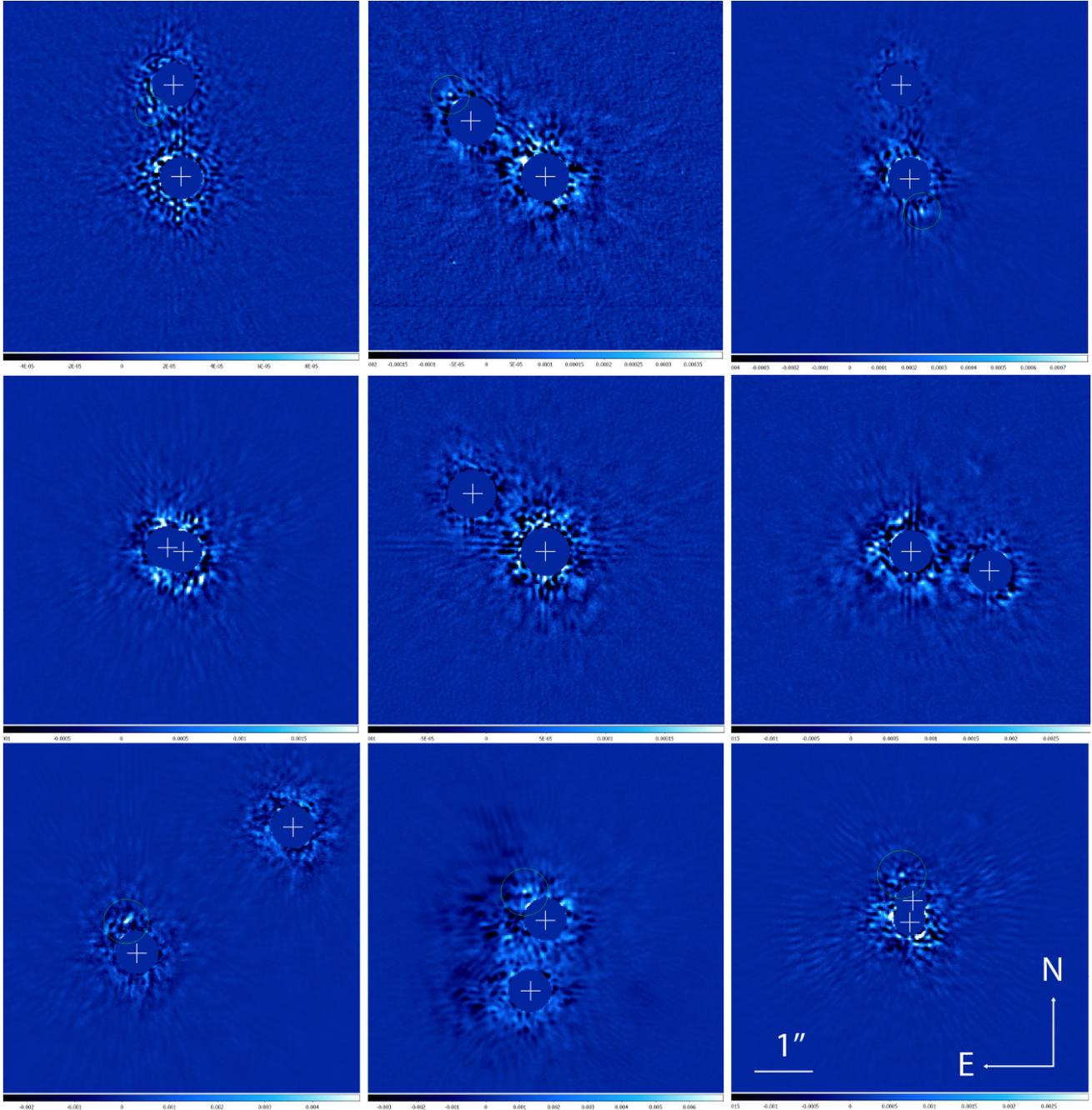

Figure 8. Mosaic of a subset of RDI-PCA reduced images, obtained on the course of our pilot survey of ~15 young binaries.

# 4. CONCLUSIONS AND PERSPECTIVES

The Hale Telescope is the only facility in the world equipped with a high-contrast instrument including a dual-stage focal-plane coronagraph like SDC, particularly downstream a 2nd-generation extreme AO system. In addition, the observation of compact binaries with two vector vortex coronagraphs in cascade fundamentally benefit from the absence of sky rotation on the 200-inch, preventing the field to rotate with respect to the instrument, hence naturally maintaining the alignment of both stars to the pair of focal-plane phase masks. We think that this may be a potential niche parameter space solely in reach of the 200-inch at the moment, although we are still only in the early phase of being able to quantify the exact rather complex boundaries of this newly-accessible discovery space, as they are deemed to depend on the actual target astrometric and brightness parameters. Although the ability to image tight observationally-challenging binaries, which most other high-contrast surveys would have a tendency to avoid, is compelling in itself, a few caveats limit the actual number of accessible young binary targets. Indeed, small-IWA phase coronagraphs like the vortex require extreme AO correction, and given the Hale Telescope equips "only" a 5-m class primary mirror, it means that binary targets with about Rmag < 8 can solely be observed under high Strehl regime. Further, given that the most young nearby moving groups are in the South hemisphere sky, it means that this intersection of brightness, age and distance requirements end up to a sample size in the order of our pilot survey (i.e. ~15 – 20 young binary stars). Once our current exploratory survey come to completion, we will investigate whether we still have a discovery space advantage – as compared to non-coronagraphic or single-coronagraph observations - under lower Strehl ratio (SR) regime, relying on the AO wavefront sensor binning modes offered by P3K,[32] which can theoretically provide SR > 0.5 for stellar magnitude down to Rmag ~ 11. This would considerably enlarge the binary sample size, and consequently represent a stepping stone towards setting-up a larger survey program at Palomar.

As for carryout of this type of dual-stage coronagraphic observations on another observatory, particularly in the South hemisphere to be able to access a larger sample, we note that we are currently not aware of any other high-contrast instrument providing two accessible intermediate focal-planes compatible with coronagraphy. Additional requirements include the availability of at least one downstream Lyot pupil-plane, remote mechanical actuation of either the intermediate image or the 2nd coronagraph along both planar coordinates, and the absence of sky rotation. The later in principle prevents the use of Alt-Az telescopes, as even the so-called "field-tracking" observing mode lets the pupil rotates, which can be detrimental to coronagraphy – unless a continuously rotating Lyot mask is procured. However, in practice this limitation needs to be evaluated, as the contrast penalty for, e.g., a support spider being not masked in the Lyot pupil-plane, is usually quite modest as compared to wavefront control aspects.[33] Another solution compatible with field rotation and ADI – and which could also address the case of higher multiple systems (triples, etc.) – would be to rely on some kind of adaptive focal-plane coronagraphs, which would null several stars in the field-of-view and "follow" their rotation while observing. Reflective liquid-crystal on silicon spatial light modulators (LCOS SLMs) are holding some promises there,[34,35] although this approach is still at a relatively low readiness level (lab demonstration in visible monochromatic light).

# ACKNOWLDGEMENTS

The authors would like to thank the Palomar Observatory day- and night-time staff for the tremendous support, particularly Rick Buruss for the initial proof-of-feasibility AO tests. JK is being funded by the Swiss National Science Foundation (SNSF) through Ambizione grant #PZ00P2_154800, and also acknowledge the Institute of Particle Physics and Astrophysics of ETH Zurich and Prof. H. M. Schmid for the observing travel support. Part of this work was carried out at the Jet Propulsion Laboratory, California Institute of Technology, under contract with the National Aeronautics and Space Administration (NASA). The data presented in this paper are based on observations obtained at the Hale Telescope, Palomar Observatory, as part of a continuing collaboration between Caltech, NASA/JPL, and Cornell University.